\documentclass[a4paper,11pt]{article}
\usepackage{pos,color,soul}

\newcommand{\beq}{\begin{equation}}
\newcommand{\eeq}{\end{equation}}

\newcommand{\sd}{\mathrm d}
\newcommand{\eexp}{\mathrm{e}}
\newcommand{\I}{\mathrm{i}}

\newcommand{\om}{\omega}

\title{Estimation of the photon production rate using imaginary momentum correlators}
\author*[a]{Csaba T\"or\"ok}
\author[b]{Marco C{\`e}}
\author[c]{Tim Harris}
\author[a]{Ardit Krasniqi}
\author[a,d,e]{Harvey B. Meyer}
\author[a]{Samuel Ruhl}

\affiliation[a]{$PRISMA^+$ Cluster of Excellence \& Institut für Kernphysik, Johannes Gutenberg-Universität Mainz,\\
  Saarstr. 21, 55122 Mainz, Germany}

\affiliation[b]{Albert Einstein Center for Fundamental Physics (AEC)
and Institut für Theoretische Physik,
Universität Bern, Sidlerstrasse 5, CH-3012 Bern, Switzerland}

\affiliation[c]{School of Physics and Astronomy, University of Edinburgh,\\
EH9 3JZ, United Kingdom}

\affiliation[d]{Helmholtz Institut Mainz, Johannes Gutenberg-Universität Mainz,
Saarstr. 21, 55122 Mainz, Germany}

\affiliation[e]{GSI Helmholtzzentrum f\"ur Schwerionenforschung,
Planckstrasse 1, 64291, Darmstadt, Germany}

\emailAdd{ctoeroek@uni-mainz.de}

\abstract
{
The thermal photon emission rate is determined by the spatially transverse, in-medium spectral
function of the electromagnetic current. Accessing the spectral function using Euclidean data is,
however, a challenging problem due to the ill-posed nature of inverting the Laplace transform. In
this contribution, we present the first results on implementing the proposal of directly computing the
analytic continuation of the retarded correlator at fixed, vanishing virtuality of the photon via the
calculation of the appropriate Euclidean correlator at imaginary spatial momentum. We employ
two dynamical O($a$)-improved Wilson fermions at a temperature of 250 MeV.
\vspace{1cm} \hspace{10.4cm}
}

\FullConference{%
	The 39th International Symposium on Lattice Field Theory, LATTICE2022
	8 Aug 2022 -- 13 August 2022
	Bonn
}

\begin{document}
\maketitle

\section{Introduction}

Ultrarelativistic heavy ion collisions have been shown to produce
a novel state of matter, the quark-gluon plasma (QGP)~\cite{PHENIX:2004vcz}.
Electromagnetic probes -- photons and dileptons -- may escape the plasma
carrying unaltered information about it, since they do not interact with
the QGP via the strong interaction.
Characterizing real photons according to their sources, we distinguish
direct and decay photons, the latter coming from the decay of final state
hadrons, while direct photons are produced during the heavy ion collision,
before the freeze-out~\cite{David:2019wpt,Gale:2021zlc}.
At low transverse momentum, the direct photon signal receives a dominant
contribution from thermal photons coming from the QGP.

Calculating the thermal photon rate of the QGP is a challenging task.
As the coupling of QCD decreases for high energies according to asymptotic freedom,
the calculation of the thermal photon rate is possible by using perturbative
methods~\cite{Jackson:2019mop,Jackson:2019yao}.
These weak-coupling results, however, become reliable only at sufficiently high
temperatures.
At strong couplings, the AdS/CFT correspondence allows for the calculation of the thermal
photon rate as well as other transport coefficients in e.g. $\mathcal{N}=4$
supersymmetric Yang--Mills theory, which shares certain common
features with QCD and is often used for comparison~\cite{Caron-Huot:2006pee}.
Using lattice QCD, one can effectively simulate QCD at strong coupling
and can also access temperatures which are close to the chiral crossover temperature.
However, lattice simulations are performed in Euclidean spacetime and the
analytic continuation of the correlation functions to Minkowskian spacetime
via an inverse Laplace transformation is a notoriously difficult, ill-posed problem
~\cite{Meyer:2011gj,Aarts:2020dda,Kaczmarek:2022ffn}.
Recent lattice QCD studies addressing the determination of the thermal photon rate
are Refs.~\cite{Ghiglieri:2016tvj,Ce:2020tmx,Ce:2022fot}.

In order to retrieve relevant information for estimating the thermal
photon rate, in this contribution we explore a novel method for the 
extraction of the thermal photon rate from Euclidean lattice QCD data
~\cite{Meyer:2018xpt}.
\section{Probing the photon rate using imaginary spatial momentum correlators}
We begin with the definition of the spectral function of the electromagnetic
current,
\beq
	\rho_{\mu\nu}(\om,{\bf k}) = \int \sd^4 x\, e^{i(\om t - {\bf k}{\bf x})}\, 
	\langle [J_\mu^{\rm{em}}(x), J_\nu^{\rm{em}}(0)^\dagger] \rangle,
\eeq
where the electromagnetic current is $J_\mu^{\rm{em}}(x) = \sum_{\rm{f}} Q_{\rm{f}} \bar{\psi}_f(x) \gamma_\mu \psi_f(x)$,
$Q_f$ being the charge of quark with flavor $f$, and the time evolution is given
in Minkowskian time by $J_\mu^{\rm{em}}(x)= e^{\I H t} J_\mu^{\rm{em}}(0) e^{-\I H t}$.
The thermal photon emission rate per unit volume of the QGP, $\sd \Gamma_\gamma(\om)/{\sd \om}$,
can be determined at leading order in the electromagnetic coupling constant as~\cite{McLerran:1984ay}:
\beq
	\frac{\sd \Gamma_\gamma(\om)}{\sd \om} = \frac{\alpha_\mathrm{em}}{\pi} \, 
	\frac{\om \sigma(\om)}{e^{\om/T}-1} + \mathcal{O}(\alpha_\mathrm{em}^2),
    \label{eq:photonrate}
\eeq
where $\sigma(\om) \equiv \rho_T(\om, k=\om)$ and $\rho_T(\om, k) = \frac{1}{2}(\delta^{ij} - k^i k^j/k^2) \rho_{ij}(\om,{\bf k})$ is the transverse channel spectral function.

\begin{figure}[b]
\begin{center}
\includegraphics[scale=0.75]{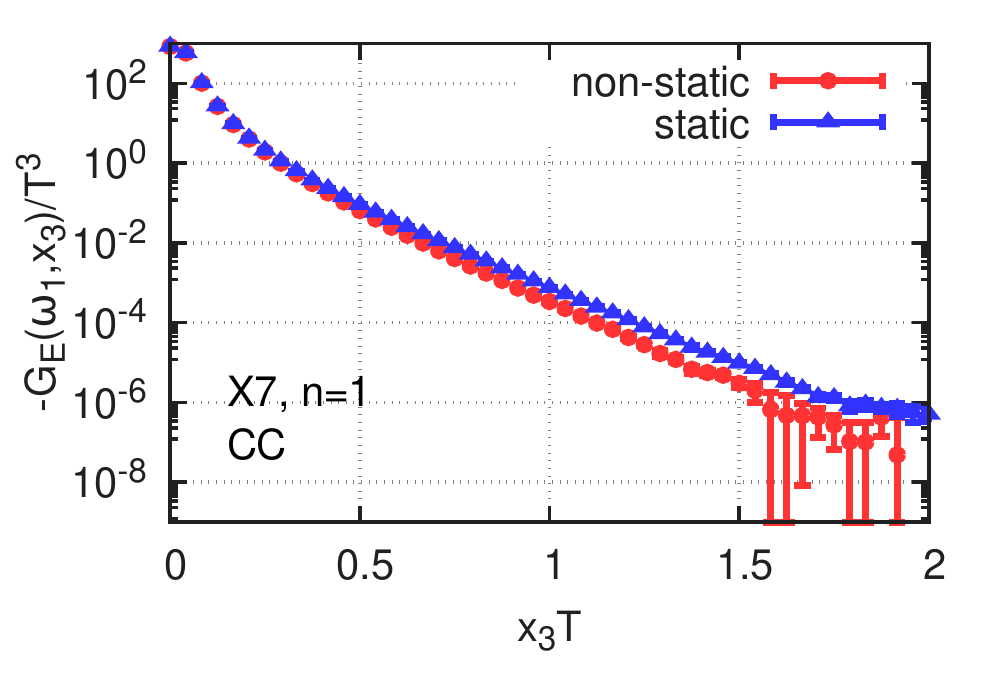}
\caption{The conserved-conserved (CC) renormalized non-static and static screening correlators
in the first Matsubara-sector on our finest ensemble called X7 ($a \sim$ 0.033 fm).}
\label{fig:ren.corr.}
\end{center}
\end{figure}

The dispersion relation which relates the spatially transverse Euclidean correlator
at imaginary momentum $k=\I \om_n$ to the spectral function at vanishing
virtuality is
\beq
	H_E(\om_n) = -\frac{\om_n^2}{\pi} \int_0^\infty \frac{\sd \om}{\om}
	\frac{\sigma(\om)}{\om^2+\om_n^2}
    \label{eq:HE,Q^2=0}
\eeq
and has been derived in Ref.~\cite{Meyer:2018xpt}.
Here, $\om_n=2 n \pi T$ is the $n$th Matsubara-frequency, and $H_E$ is the
Euclidean transverse channel current-current correlator evaluated at imaginary
spatial momentum,
\begin{align}
	H_E(\om_n) \equiv G_E^T(\om_n,k={i \om_n}) =
	- \int_0^\beta \sd x_0 \int \sd^3 x\,
	\eexp^{\I \om_n x_0}\, \eexp^{\om_n x_3} \langle J_1(x) J_1(0) \rangle.
	\label{eq:HE,def}
\end{align}
The short-distance convergence properties of this correlator can be analyzed
by expanding $e^{\I \om_n x_0}$ and recalling that the short-distance behavior
of the current-current correlator starts with $1/x^6$.
Within the continuum theory, the correlator vanishes in the vacuum, but this
property is lost at finite lattice spacing due to the lack of Lorentz symmetry.
The property can be restored by subtracting a correlator with the same short-distance
behavior and -- in order to not to alter the continuum limit --, that vanishes in
the continuum.
One can achieve this either by subtracting the vacuum lattice correlator obtained
at the same bare parameters~\cite{Meyer:2018xpt} or by subtracting a thermal
lattice correlator having the same momentum inserted into a spatial direction~\cite{Meyer:2021jjr}.
Since the latter option does not require additional simulations at $T=0$, we
proceed using the following estimator
\begin{align}
	H_{E}^{\rm{(sub)}}(\om_n) &= - \int_0^\beta \sd x_0 \int \sd^3 x\,
	\left( {\eexp^{\I \om_n x_0}} - {\eexp^{\I \om_n x_2}} \right)\,
	\eexp^{\om_n x_3} \langle J_1(x) J_1(0) \rangle \nonumber \\
	&= - \int_{-\infty}^{\,\infty} \sd x_3\, \eexp^{\om_n x_3}
	\Big[ {G_{\rm{s}}(\om_n,x_3)} - {G_{\rm{ns}}(\om_n,x_3)} \Big],
\label{eq:HEsub,def}
\end{align}
where in the second line we introduced the static ($G_{\rm{s}}$) and non-static ($G_{\rm{ns}}$)
transverse screening correlators, involving $e^{\I \om_n x_2}$ and $e^{\I \om_n x_0}$,
respectively.
By performing the subtraction as in Eq.~(\ref{eq:HEsub,def}), the contribution of the unit
operator cancels and the resulting expression is integrable.
Moreover, the estimator given in Eq.~(\ref{eq:HEsub,def}) vanishes in the vacuum.
By inserting the imaginary as well as the real spatial momentum to different combinations
of directions and then averaging, we increased the statistics for the evaluation
of this observable.
The static and non-static screening correlators are shown in Fig.~\ref{fig:ren.corr.}.
In the following we omit the upper index '(sub)' referring to 'subtracted' from our
lattice estimator.

\section{Lattice setup}
In order to calculate the screening correlators which enter the expression~(\ref{eq:HEsub,def})
for $H_E$, we used three ensembles generated at the same temperature in the high-temperature phase
($T \sim 250$ MeV).
We employ two-flavors, O($a$)-improved dynamical Wilson fermions and the plaquette gauge action.
The zero-temperature pion mass in our study is around $m_\pi \approx 270$ MeV, and the lattice 
spacings are in the range of 0.033--0.05 fm.
We use the isovector vector current instead of the electromagnetic current,
whereby disconnected contributions are absent.
The Euclidean correlators at imaginary spatial momentum have been measured using the
local as well as the conserved discretizations of the currents both at source
and sink, resulting in total of four different discretizations (local-local,
conserved-conserved, local-conserved and conserved-local).
The two mixed discretizations are not independent, they can be transformed 
into each other using Cartesian coordinate reflections.
After averaging these appropriately, we had therefore three different
discretizations of the correlators.
At each ensemble we had around 1500--2000 configurations and 64 point sources
per configuration.
We renormalized the local-local and the mixed correlators by multiplying by $Z_V^2$
or $Z_V$, respectively.
We took the corresponding value for $Z_V$ from Ref.~\cite{DallaBrida:2018tpn}.

\begin{figure}[b]
\begin{center}
\includegraphics[scale=0.72]{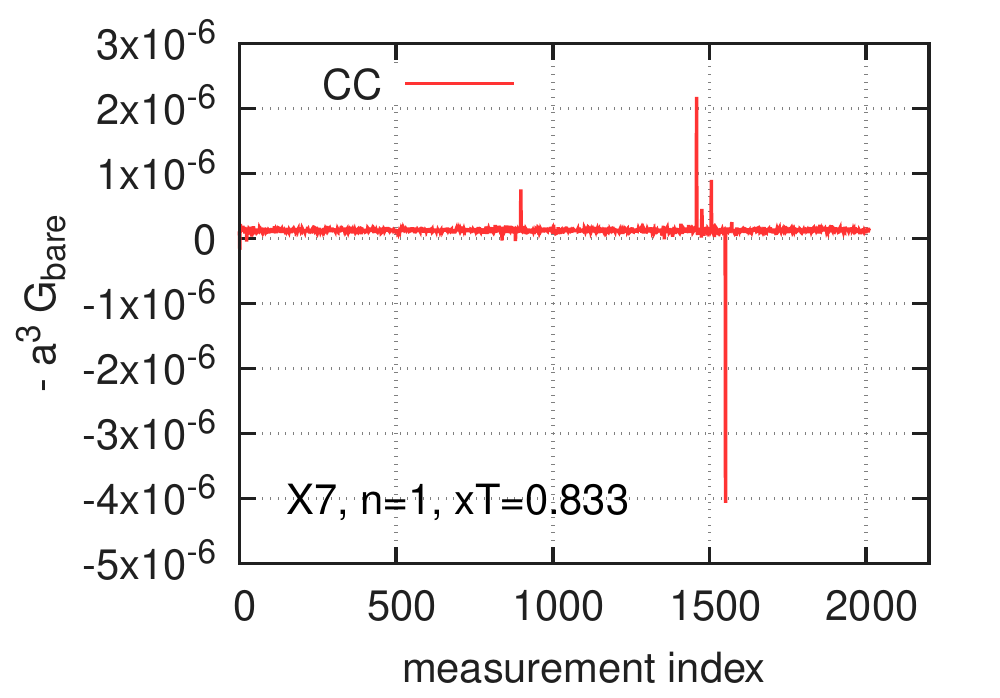}
\includegraphics[scale=0.75]{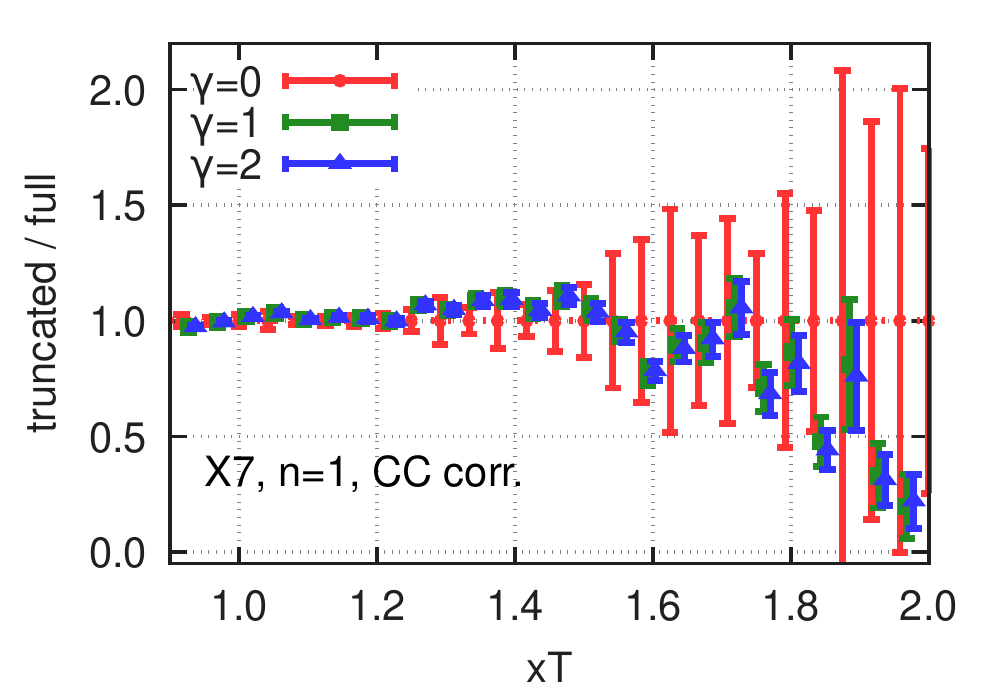}
\caption{Left panel: Measurement history of the non-static screening correlator at $xT=0.833$.
The outliers are shown up as spikes in the data.
Right panel: Truncation stability for the non-static screening correlator.
The data points corresponding to the trimmed data have been shifted slightly
to the right to improve visibility.}
\label{fig:outliers&truncation}
\end{center}
\end{figure}

We occasionally encountered results for the correlator at a certain Euclidean
distance with several standard deviations off from the mean value at that distance,
which we identified as outliers, see the left panel of Fig.~\ref{fig:outliers&truncation}.
These occured more frequently at larger Euclidean separations.
These outliers increased the statistical error and also modified the mean to some extent,
see Fig.~\ref{fig:outliers&truncation}, right panel.
We eliminated these outliers by using robust statistics~\cite{Huber:2009snd}.
First, we prepared a distribution of results at each Euclidean distance, then removed
the data points belonging to the lower and upper $\gamma$\% of that distribution.
We varied $\gamma$ between 0.5--4 when making these cuts and found that the
error estimation as well as the calculation of the mean is more stable this way.
In the final analysis we used $\gamma=1$, but when we detected only less than 10 datapoints
being outside five times the interquantile range from the mean, we only applied
trimming with $\gamma=0.5$.
We show an example in case of the conserved-conserved correlator at our finest
ensemble, X7, in the right panel of Fig.~\ref{fig:outliers&truncation}.
At short distances of the correlator, this approach did not influence the
results, because outliers occured there only very rarely.
At intermediate distances, i.e. around $xT\sim$0.7--1.3, the effect of this
method was again not significant.
At large distances, however, the errors reduced by a factor of around 2--6
when omitting the tails of the distributions.
Since we believe that these outliers are not likely to have physical
origin, their exclusion should not influence the validity of the extracted
physical results.
This is indeed what we found when analyzing the data without truncating:
we obtained consistent final results but with larger errors.
\section{Modeling the tail of the screening correlators}

\begin{figure}[b]
\begin{center}
\includegraphics[scale=0.75]{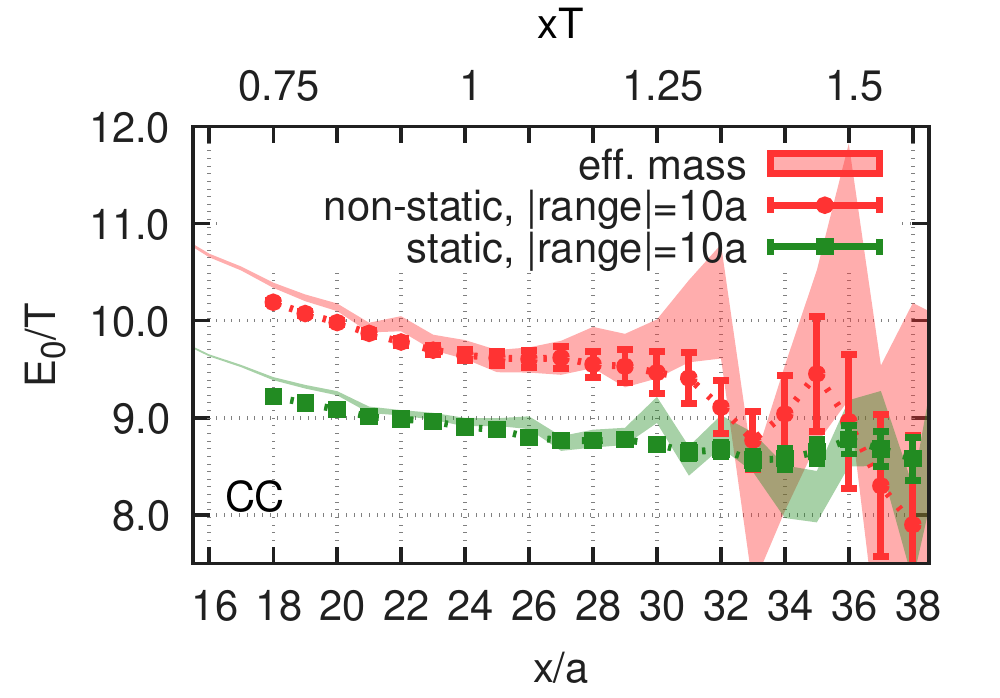}
\includegraphics[scale=0.68]{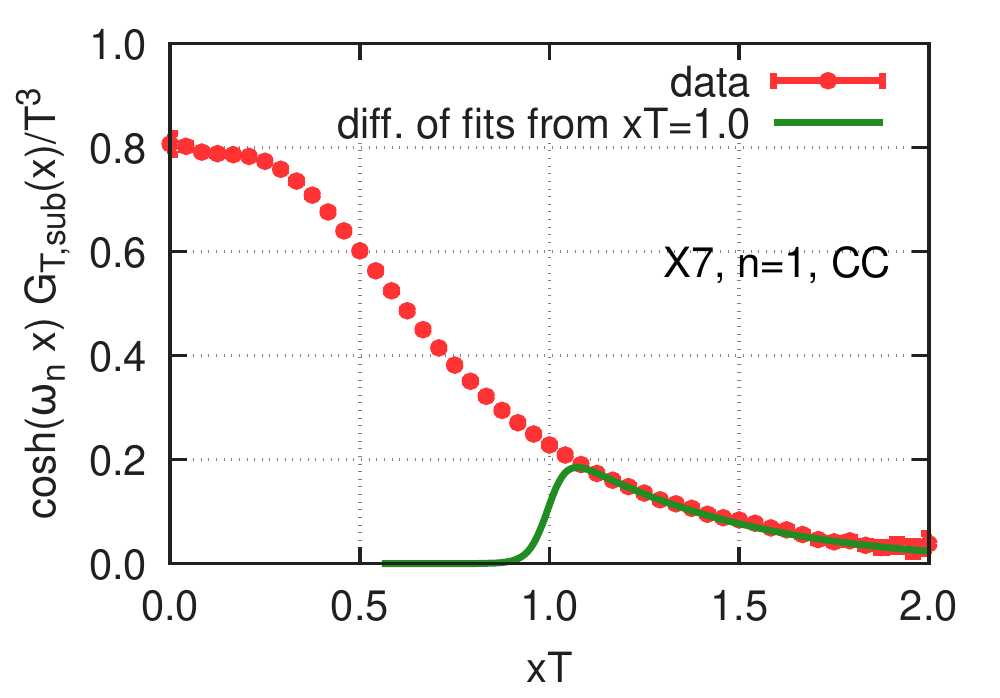}
\vspace{-0.2cm}
\caption{Left: fitted masses and effective masses for the non-static and static
screening correlators using a fit range of 10$a$.
Right: the integrand needed for the calculation of $H_E$ at the first Matsubara-frequency.}
\label{fig:extrmass&intnd}
\end{center}
\end{figure}

Since the integral of Eq.~(\ref{eq:HEsub,def}) receives contributions from large
distances as well, we need to get good control over the screening correlators at
large distances, if we aim at a precise determination of $H_E$.

The screening correlators have a representation in terms of energies and amplitudes
of screening states in the following form~\cite{Meyer:2018xpt}:
\beq
	G_{\rm{ns}}(\om_r,x_3) \overset{x_3 \ne 0}{=}
	\sum_{n=0}^{\infty} |A_{\rm{ns},n}^{(r)}|^2 e^{-E_{\rm{ns},n}^{(r)} |x_3|}.
\eeq
A similar expression holds for the static correlator.
The low-lying screening spectrum can be studied using weak-coupling methods
as well~\cite{Brandt:2014uda}.
The lowest energy of a screening state in a given Matsubara sector with frequency
$\om_r$ is often called the screening mass and is denoted by $E^{(r)}_0$.

In order to get a better handle on the asymptotic behavior of the screening
correlators and avoid the enhancement of the error on $H_E$ coming from fluctuations
present in the actual correlators at large distances, we performed single-state
fits on the tails of the correlators using the above representation translated
to a form corresponding to a periodic lattice, namely
\beq
	G_{\rm{ansatz}}(\om_r,x_3) = |A_0^{(r)}|^2 \cosh\big[E^{(r)}_0 (x_3-L/2)\big],
\eeq
$L$ being the spatial length of the lattice.
While the single-state fits describe the actual data well, i.e. with good $\chi^2$-
and p-values, the identification of the plateau region was not clear in several
cases, although we performed a thorough scan using all possible fit ranges having
different starting points and different lengths with 6$a$--11$a$.
Besides fitting, we also determined the "effective mass" using two consecutive
correlator datapoints, by solving
\beq
	\frac{G(\om_r,x_3+a)}{G(\om_r,x_3)}
	= \frac{\cosh\big[m_{\rm{eff}} (x_3+a-L/2)\big]}{\cosh\big[m_{\rm{eff}} (x_3-L/2)\big]}
\eeq
for $m_{\rm{eff}}$.
The effective masses are in quite good agreement with the fitted masses, but also
do not show a clear plateau as $x_3$ increases, see Fig.~\ref{fig:extrmass&intnd}, left panel.
Therefore, we decided to choose three representatives from a histogram built by
assigning Akaike-weights~\cite{Akaike:1973abc,Borsanyi:2020mff} to all the fitted
masses that we obtained.
We propagate the median as well as the values near the 16th and 84th percentiles
to the later steps of the analysis.
When proceeding this way for the non-static as well as for the static screening
correlators, we obtain $3\times 3=9$ possibilities for modeling the tail of
the integrand of Eq.~(\ref{eq:HEsub,def}) on a given ensemble.
We calculated $H_E$ using all these nine combinations for the tail, sorted the
results and then chose the median, the values near the 16th and near
the 84th percentile after assigning uniform weights for these slightly different
values of $H_E$.
Thus for each ensemble we had three representative values of $H_E$ that went into
the next step of the analysis, which was the continuum extrapolation.
We note that by modeling the tail of the non-static and static screening correlators
by doing single-state fits, we could reduce the errors by a factor of around 2.5
on our coarsest ensemble.
The transition to the modelled tail has been introduced smoothly by using
a step function and we investigated the effect of choosing different switching
points, $x_w$, in the range $x_w T$=0.8--1.2.
We found that the results were essentially stable against these choices.
\section{Continuum extrapolation}
\begin{figure}[b]
\begin{center}
\includegraphics[scale=0.74]{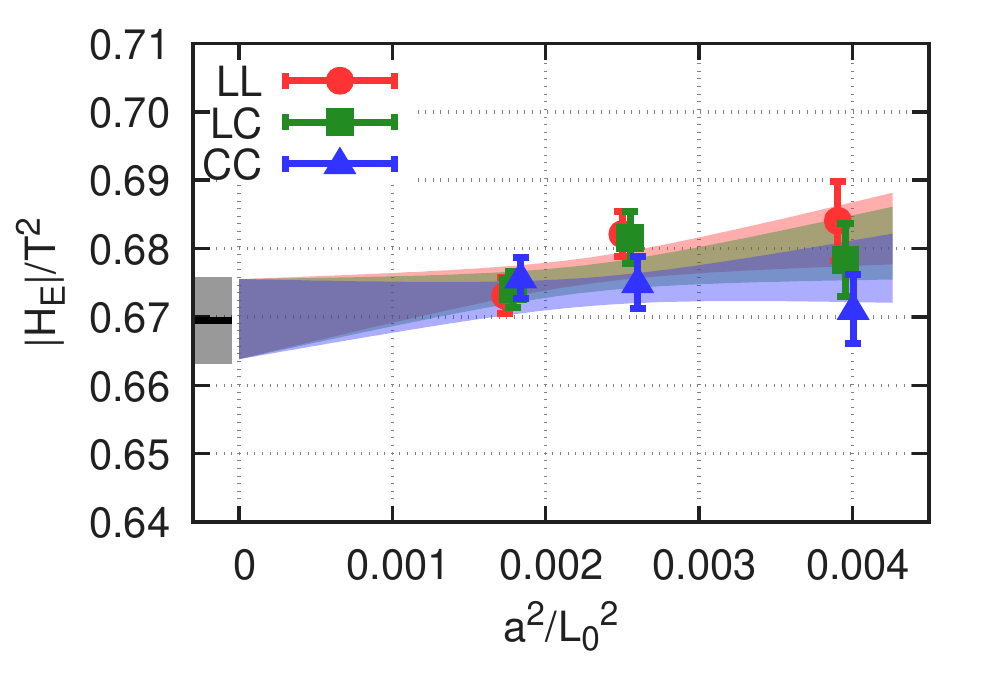}
\includegraphics[scale=0.74]{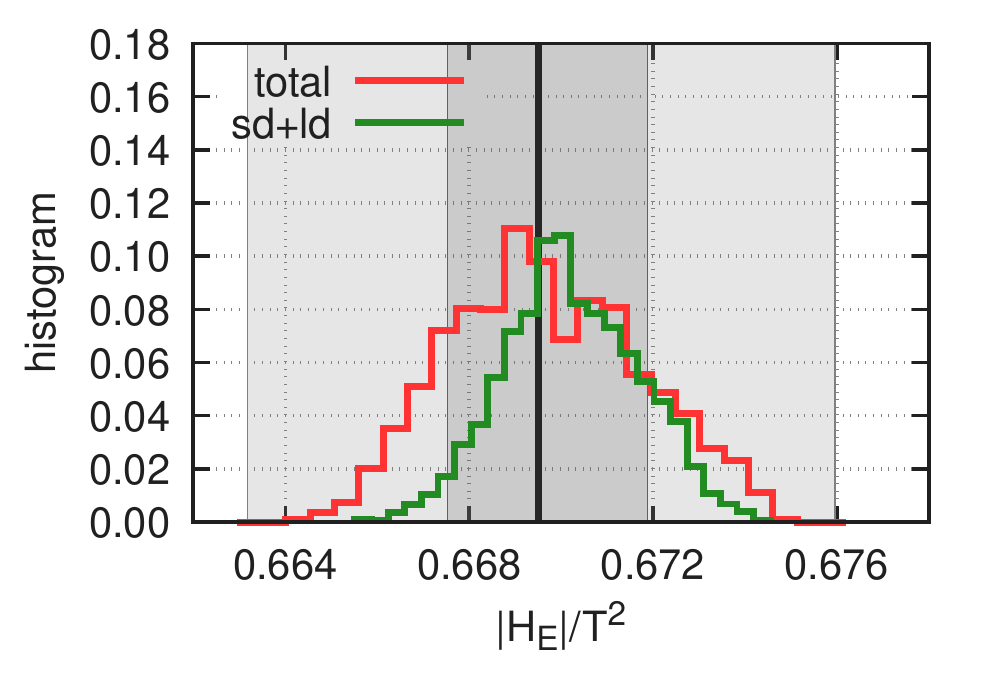}
\caption{Left: a representative continuum limit of $H_E$.
Right: histogram of the Akaike-weights plotted against the continuum 
extrapolated results.
The AIC-weighted histogram of the long-distance contribution to $H_E$ shifted
with the continuum result for the short-distance contribution is also shown for 
comparison (sd+ld) and are in agreement with the AIC-weighted histogram of the 
continuum extrapolated results of the total $H_E$.
}
\label{fig:contlim}
\end{center}
\end{figure}

The three discretized correlators allowed us to perform a correlated simultaneous
continuum extrapolation of $H_E$.
We used a linear ansatz in $a^2$ when extrapolating to the continuum.

As discussed in the previous section, in order to have a more precise value
we modelled the tail of the integrand needed to evaluate $H_E$.
Since this way at each ensemble and for each discretization we had three
representative values of $H_E$, we used these in all possible combinations when
performing the continuum limit.
These gave a total of $(3^3)^3=19683$ different continuum extrapolations of which
we built an AIC-weighted histogram to estimate the systematic error.
A representative continuum extrapolation as well as the histogram
are shown in Fig.~\ref{fig:contlim}, left and right panel, respectively.

We also performed separate continuum extrapolations of the short- as well as
of the long-distance contributions.
In the case of the short-distance contribution, we simply integrated using the
trapezoid formula and the systematic error of the continuum extrapolation has been
estimated by omitting one of the discretizations at the coarsest ensemble.
By shifting the Akaike-weighted histogram of the continuum extrapolated values of
the long-distance contribution with the central value of the continuum extrapolated
short-distance contribution, we observe that it is consistent with the 
continuum extrapolated values of the total $H_E$ (Fig.~\ref{fig:contlim}, right panel).

\vspace{-0.4cm}

\section{Comparisons}
\begin{table}[t]
\begin{center}
\begin{tabular}{ll}
\hline
\hline
observable	& value at the continuum limit \\ \hline
$|H_E|/T^2$	& \quad $0.670(6)_{\rm stat}(2)_{\rm sys}$ \\
sd($x_{\rm w}T=1)$ 	& \quad $0.579(3)_{\rm stat}(1)_{\rm sys}$ \\
ld($x_{\rm w}T=1)$ 	& \quad $0.091(5)_{\rm stat}(2)_{\rm sys}$ \\
\hline
\hline
\end{tabular}
\end{center}
\vspace{-0.3cm}
\caption{Results in the first Matsubara sector using single-state fits to model
the tail of the integrand.}
\end{table}

For the purpose of comparison, it is worth calculating the imaginary
part of the retarded correlator at the lightcone in the free theory
as well as in strongly coupled $\mathcal{N}=4$ super Yang--Mills theory
using the AdS/CFT correspondence.
This has been done in Ref.~\cite{Meyer:2018xpt}.
In the free theory, $|H_E|/T^2 = 0.5$ in the first Matsubara sector, while
in $\mathcal{N}=4$ super Yang--Mills theory, $|H_E|/T^2 \approx 0.75$.
When normalizing with the temperature, the lattice result we obtained,
$0.670(6)_{\rm{stat}}(2)_{\rm{sys}}$, is between these two values.
Using another normalization might also be interesting.
When dividing by the static susceptibility, $\chi_s$, of the 
relevant theories, we obtain in the first Matsubara sector the following
results: $[|H_E|/\chi_s]^{\rm{(free)}}= 0.5$ in the free theory (because $\chi_s^{\rm{(free)}}/T^2=1$),
$[|H_E|/\chi_s]^{\rm{(SYM)}} \approx 0.67$ in $\mathcal{N}=4$ super Yang--Mills theory
with $N_c=3$.
On the lattice, we determined the static susceptibility in Ref.~\cite{Ce:2022fot}
to be $\chi_s^{\rm{(lat)}}/T^2=0.882(11)_{\rm{stat}}(19)_{\rm{sys}}$.
Using this normalization, the lattice result is $[|H_E|/\chi_s]^{\rm{(lat)}} \approx 0.76$
that is about 13\% larger than the value in $\mathcal{N}=4$ SYM theory.
\section{Conclusions and outlook}
In this contribution, we calculated Euclidean correlators at imaginary
spatial momentum, that are related to the thermal photon emission rate
according to Eq.~(\ref{eq:HE,Q^2=0}).
We focused on the correlator evaluated at the first non-vanishing Matsubara frequency.
In order to improve the predictive power of our result we modelled the tail of the
screening correlators occuring in the integrand for our primary quantity $H_E$.
We were able to describe the data with single-state fits with good p-values.
We performed a simultaneaous correlated continuum extrapolation of the
three lattice discretizations of the imaginary momentum correlator
using three thermal ensembles.
The result we obtained is in the same ballpark as obained in the free
theory or in $\mathcal{N}=4$ supersymmetric Yang--Mills theory.
Depending on the normalization it could be between these two, or 
larger than these results.

As noted in Ref.~\cite{Meyer:2018xpt}, the knowledge of $H_E(\om_n)$ for all $n>n_0$
would enable one to determine the spectral function uniquely by
Carlson's theorem.
Following a similar route for analyzing $H_E$ in the second Matsubara-sector,
however, revealed that the signal will soon become noisy and the uncertainty
of $H_E(\om_{n=2})$ is therefore much larger.
At present our determination of $H_E(\om_{n=2})$ is compatible within errors
with the $H_E(\om_{n=1})$ result.
We note, however, that since $\sigma(\om)>0$, $|H_E(\om_r)|$ has to be larger
than $|H_E(\om_n)|$ if $r>n$ ~\cite{Meyer:2018xpt,Meyer:2021jjr}.
\footnote{
	Note however that without taking the absolute value, the ordering is $H_E(\om_r) < H_E(\om_n)$,
	when $\om_r>\om_n$, since $H_E$ is negative.
}
In order to have a reliable calculation of $H_E(\om_{n \ge 2})$, one has
to implement algorithmic improvements and/or devise other operators which
could help to constrain the long-distance behavior of the screening
correlators.
The exploration of these directions is left for future work.
\vspace{-0.1cm}
\section{Acknowledgements}
\vspace{-0.1cm}

This work was supported by the European Research Council (ERC) under the European
Union’s Horizon 2020 research and innovation program through Grant Agreement
No.\ 771971-SIMDAMA, as well as by the Deutsche Forschungsgemeinschaft 
(DFG, German Research Foundation) through the Cluster of Excellence “Precision Physics,
Fundamental Interactions and Structure of Matter” (PRISMA+ EXC 2118/1) funded by
the DFG within the German Excellence strategy (Project ID 39083149).
T.H. is supported by UK STFC CG ST/P000630/1.
The generation of gauge configurations as well as the computation of correlators was
performed on the Clover and Himster2 platforms at Helmholtz-Institut Mainz and on Mogon II
at Johannes Gutenberg University Mainz.
We have also benefitted from computing resources at Forschungszentrum J\"ulich allocated
under NIC project HMZ21.
For generating the configurations and performing measurements, we used the openQCD~\cite{Luscher:2012av}
as well as the QDP++ packages~\cite{Edwards:2004sx}, respectively.


\setlength{\bibsep}{0pt plus 0.3ex}

\end{document}